\documentclass[letterpaper, onecolumn,12pt]{IEEEtran}

\usepackage{setspace}
\usepackage{fancyhdr}
\usepackage{graphicx}
\usepackage{url}
\usepackage{multirow}
\usepackage{color}
\usepackage{placeins}
\usepackage{float}
\usepackage{tabularx,colortbl}
\usepackage{amsmath}
\usepackage{amsfonts}
\usepackage{amssymb}
\usepackage{amsthm}
\usepackage{epstopdf}
\usepackage{cite}
\usepackage{amsmath,amssymb,amsfonts}
\usepackage{bm}
\usepackage{algorithmic}
\usepackage{graphicx}
\usepackage{textcomp}
\usepackage{xcolor}
\usepackage{amsfonts}
\usepackage{bm}
\usepackage{bm}
\usepackage[square, comma, sort&compress, numbers]{natbib}

\begin{document}
%
\title{Bit-level Optimized Neural Network for Multi-antenna Channel Quantization}
\author{
\IEEEauthorblockN{Chao Lu\IEEEauthorrefmark{0}, Wei Xu\IEEEauthorrefmark{0}, Shi Jin\IEEEauthorrefmark{0}, and Kezhi Wang\IEEEauthorrefmark{0}}

\vspace{-1cm}

\thanks{Manuscript received June 17, 2019; accepted September 15, 2018. This work was supported in part by the National Key Research and Development Program 2018YFA0701602, the Natural Science Foundation of Jiangsu Province for Distinguished Young Scholars under SBK2019010264, the NSFC under grants 61871109, and the Royal Academy of Engineering under the Distinguished Visiting Fellowship scheme. The editor coordinating the review of this paper and approving it for publication was Prof. J. Mietzner. \emph{(Corresponding author: Wei Xu)}

C. Lu, W. Xu and S. Jin are with the National Mobile Communications Research Laboratory, Southeast University, Nanjing 210096, China (\{220170709, wxu, jinshi\}@seu.edu.cn).

K. Wang is with the Department of Computer and Information Sciences, Northumbria University, Newcastle upon Tyne, NE1 8ST, UK (kezhi.wang@northumbria.ac.uk).
}
}

\maketitle

\newtheorem{mylemma}{Lemma}
\newtheorem{mytheorem}{Theorem}
\newtheorem{mypro}{Proposition}
\begin{abstract}
Quantized channel state information (CSI) plays a critical role in precoding design which helps reap the merits of multiple-input multiple-output (MIMO) technology. In order to reduce the overhead of CSI feedback, we propose a deep learning based CSI quantization method by developing a joint convolutional residual network (JC-ResNet) which benefits MIMO channel feature extraction and recovery from the perspective of bit-level quantization performance. Experiments show that our proposed method substantially improves the performance.

\end{abstract}

\begin{IEEEkeywords}
Channel state information (CSI), quantization, neural network (NN), multiple-input multiple-output (MIMO).

\vspace{-0.5cm}
\end{IEEEkeywords}

\IEEEpeerreviewmaketitle
\section{Introduction}

Multiple-input multiple-output (MIMO) technology has shown its ability in obtaining rich diversity and multiplexing gains. With recent development of MIMO, the trend of deploying more antennas comes with a number of challenges. One of the major challenges is to acquire accurate high-dimensional MIMO channel state information (CSI) at the transmitter, especially in frequency division duplex (FDD) systems. An increasing number of antennas results in a high-dimensional channel matrix, which makes it difficult for the transmitter to obtain accurate CSI through a feedback channel of limited bandwidth.

Conventional codebook based methods quantize the channel matrix into a sequence of bits which represents the index of a codeword \cite{AoD}. Considering that the complexity of the codebook based methods grows exponentially with the number of quantization bits, it can be prohibitively difficult for large MIMO channels. Compressed sensing (CS) based methods, e.g., TVAL3 \cite{Tval3}, can be used for solving this difficulty through feature extraction and dimension compression. However, when the compression ratio is extremely high, the performance of these CS based methods degrades severely \cite{CsiNet}.

Recently, neural network (NN) based methods, inspired from CS based methods, showed appealing performance for MIMO CSI compression \cite{CsiNet, CsiNetLSTM1, CsiNetLSTM2}. Previously, encoder-decoder schemes were proposed in \cite{encoder_decoder1,encoder_decoder2} to learn the behaviors of a transmitter in wireless communication systems. This idea of an encoder-decoder network was then utilized to learn the MIMO CSI compression \cite{CsiNet, CsiNetLSTM1, CsiNetLSTM2}. In particular, the encoder network compressed the channel matrix into a low dimensional vector while the decoder network was responsible for recovering the channel matrix directly from the compressed vector. Usually the decoder was designed with a more complex structure than the encoder because the base station (BS) generally has stronger computing ability than user equipment (UE).

The NN based method in \cite{CsiNet}, namely CsiNet, assumed that the decoder achieves a compressed CSI vector of continuous values. Then in \cite{CsiNetLSTM1,CsiNetLSTM2}, the CsiNet was enhanced by utilizing a long-short time memory (LSTM) network \cite{LSTM} to further exploit temporal channel correlations. In a digital communication system with limited bandwidth, however, it is still impossible to transmit a vector of continuous values even though it is highly compressed. In most applications, CSI quantization is necessary and inevitable for implementational tractability.


In this work, we consider the CSI compression as well as quantization through a bit-level optimized NN.
The main contributions of this work are summarized as follows:

\begin{itemize}
\item  We propose a bit-level optimized design of NNs by considering the effects of quantization distortion. This design can be easily assembled in existing CSI networks with a few additional modifications.
\item We develop a joint convolutional residual network (JC-ResNet) to extract channel features for CSI compression. The proposed JC-ResNet is much more efficient than common convolution operations in existing NNs.
\end{itemize}

\section{CSI Quantization and Feedback}

\begin{figure*}[thbp]
\centerline{\includegraphics[width=17cm,height=3.2cm]{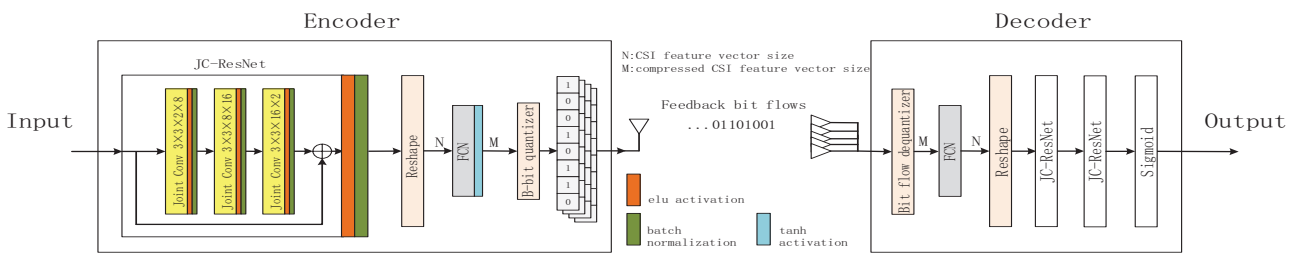}}
\caption{Architecture of a bit-level optimized network.}
\label{quant_net}
\end{figure*}

We consider an orthogonal frequency division multiplexing (OFDM) system
where the BS is equipped with $N_t$ antennas and the UE has a single antenna. Assume that the signal obtained at the $n$-th subcarrier is \begin{equation}\label{eq:ReveivedSignal}
y_n={{\bf{h}}^H_n}{{\bf{v}}_n}x_n+n_n,
\end{equation}
where ${\bf{h}}_n \in {\mathbb{C}}^{N_t \times 1} , {{\bf{v}}_n} \in {\mathbb{C}}^{N_t \times 1}, x_n \in {\mathbb{C}}$, and $n_n \in {\mathbb{C}}$ respectively represent the channel response, precoding vector, transmitting signal, and additive noise at the $n$-th subcarrier. The precoding vector ${\bf{v}}_n$ is designed according to CSI feedback. We stack the channel vector and denote ${\bf{H}}=\left[{\bf{h}}_1 {\bf{h}}_2 ... {\bf{h}}_{N_c}\right]^H  \in {\mathbb{C}}^{N_c \times N_t}$, where $N_c$ is the number of subcarriers of the OFDM system.

In this paper, we propose an encoder-decoder network to conduct CSI compression/uncompression procedures for feedback. 
Assuming that ${\bf{H}}$ is known at the UE, a general representation of the CSI quantization and feedback process can be written as: 
\begin{equation}\label{eq:quant_feedback}
{\hat{\bf{H}}} = f_d({\mathcal{D}}({\mathcal{Q}}(f_e({\bf{H}})))),
\end{equation}
where $\hat{\bf{H}}$ denotes the recovered CSI at the BS, $f_e(\cdot)$ represents the preprocessing and encoding before quantization, ${\mathcal{Q}}(\cdot)$ is the quantization operator, ${\mathcal{D}}(\cdot)$ conducts the corresponding dequantization, and $f_d(\cdot)$ represents the decoding and postprocessing function after dequantization, e.g., an inverse operation of $f_e(\cdot)$. 

\begin{figure}[t!]
\centerline{\includegraphics[width=7.9cm,height=4.2cm]{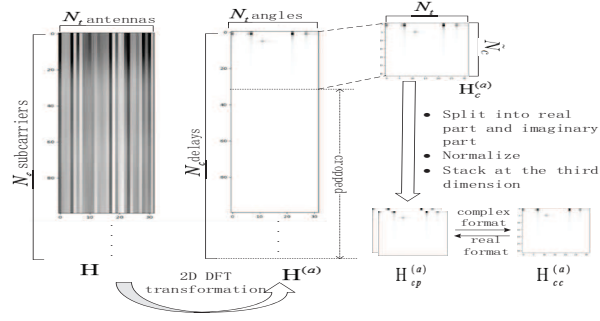}}
\caption{Pseudo graphics of preprocessed channel before network processing.}
\label{channel}
\end{figure}

A diagram of our proposed NN structure is depicted in Fig.~\ref{quant_net}.
In the proposed scheme, $f_e(\cdot)$ includes the steps as depicted in Fig. \ref{channel}. First, the raw channel matrix $\bf{H}$ acquired at the UE is transformed into the angular-delay domain through a 2D Fourier transformation. Letting ${\bf{H}}^{(a)} \in  {\mathbb{C}}^{N_c \times N_t} $ be the transformed channel matrix in the angular-delay domain, we have
\begin{equation}
 {\bf{H}}^{(a)} = {\bf{F}}_a{\bf{H}}{\bf{F}}_b^H,
\end{equation}
where ${\bf{F}}_a \in  {\mathbb{C}}^{N_c \times N_c}$ and ${\bf{F}}_b \in {\mathbb{C}}^{N_t \times N_t}$ are Fourier matrices. Considering that the maximum multipath delay is limited in practice, it implies that the entries at large delay positions are all zeros \cite{CsiNet}. Therefore, without loss of generality, we can focus on a submatrix of ${\bf{H}}^{(a)}$, i.e., keeping the first ${\tilde{N}}_c$ rows with nonzero entries at the delay domain, and get a cropped channel matrix ${\bf{H}}^{(a)}_{c} \in  {\mathbb{C}}^{{\tilde{N}}_c \times N_t}$.
If some inignorably small negative delays exist, a reshuffling of  ${\bf{H}}^{(a)}$ can be added in this scheme before the cropping. Note that this additional operation of reshuffling only generates a different version of network input, we do not need any other further changes for the proposed NN.
Note that for convenience of the processing in the NN, we then split ${\bf{H}}^{(a)}_{c}$ into two parts, i.e., real and imaginary parts, and stack the two parts at the third dimension.
Then, all the entries of the processed channel matrix are normalized to (0, 1). We denote it by ${\bf{H}}^{(a)}_{cp} \in  {\mathbb{R}}^{{\tilde{N}}_c \times N_t \times 2}$  and its corresponding complex version by ${\bf{H}}^{(a)}_{cc} \in  {\mathbb{C}}^{{\tilde{N}}_c \times N_t}$. We stress that the operation of CSI scaling does not need to be de-scaled at the recovery side because the directional information of channel is (typically) sufficient for beamforming design at the transmitter side.

Before channel quantization in the NN, our encoder network generates the compressed CSI feature vector from ${\bf{H}}^{(a)}_{cp}$.
${\mathcal{Q}}(\cdot)$ rounds the entries of the compressed CSI feature vector generated by the encoder network into their nearest integers.
Since the rounded numbers can be expressed by a binary sequence, this binary bit flow is the quantization output which can be directly fed back to the BS.
At the other side, ${\mathcal{D}}(\cdot)$ conducts the inverse operation of ${\mathcal{Q}}(\cdot)$. The decoder network recovers an estimation of channel matrix ${\bf{H}}^{(a)}_{cp}$. Then we can get $\hat{\bf{H}}$, which is the estimation of the raw channel matrix ${\bf{H}}$ from ${\bf{H}}^{(a)}_{cp}$ through an inverse transformation of the preprocessing done in $f_e(\cdot)$. These two steps make up $f_d(\cdot)$

\section{Proposed Scheme}
In this section, we describe the proposed network architecture and then elaborate the computation process of the JC-ResNet in detail.

\subsection{Network Architecture}

Encoder and decoder networks using deep NN are utilized for CSI quantization and recovery. The input of the encoder network is a cropped and normalized channel matrix in the angle-delay domain, i.e., ${\bf{H}}^{(a)}_{cp}$.
The existing encoder-decoder design works at the condition that the number of the quantization bits of the compressed vector needs to be so large that the quantized CSI vector does not have much distortion, because the compression is designed without considering quantization.

In order to optimize the CSI compression jointly with quantization, we propose the bit-level encoder-decoder architecture in Fig. \ref{quant_net}.
The encoder network encompasses the newly designed JC-ResNet. It better exploits the features of the channel matrix, and the output of this JC-ResNet includes abstract sparse features. A nonlinear ``$elu$'' activation layer and a batch normalization layer are appended to the JC-ResNet. Specifically,  the ``$elu$'' function is defined as follows:
\begin{equation}
 elu(x) = \left\{
             \begin{array}{lr}
             x, & x>0  \\
             e^x - 1, & x\le0.
             \end{array}
\right.
\end{equation}
Then, the output features are reshaped to a 1D tensor for further processing. The 1D feature tensor is compressed by a fully connected network (FCN), which is expected to help discard much redundant information and produce dense features.  Following the FCN layer, a ``$tanh$'' activation layer is added to map the output entries within $(-1,1)$.

To optimize the performance, a quantization distortion metric is considered for network training. Specifically, the low-dimensional features are quantized by a $B$-bit quantizer.
The quantization computation takes the following form:
\begin{equation}\label{eq:quant}
y = {\text{round}}{(2^{B-1} \times x)}/2^{B-1},
\end{equation}
where $B$ denotes the number of quantization bits for each entry of the compressed CSI vector, ${x}$ denotes the entries of the compressed CSI vector after the FCN, and ${\text{round}}(\cdot)$ is the rounding function. In (5), $x$ is firstly zoomed to $(-2^{B-1},2^{B-1})$ and then is rounded to its nearest integer.
In order to accomplish the quantization with $B$ bits, we alternatively use $-2^{B-1}+1$ to represent the quantization of the values whose nearest integer is $-2^{B-1}$.
Finally, the output is reverted to $(-1,1)$ by dividing by $2^{B-1}$. In this way, each entry of the compressed CSI vector has a corresponding $B$-bit expression.

The most challenging issue is that the gradient of the quantization operation is always zero except for some specific points, which is fatal to network training. According to \cite{image_compress}, the quantization operation in the network can be seen as a transparent module, which means that the gradients after the quantization are back propagated through it in a lossless fashion. In other words, we can alternatively set the quantization gradients to 1 everywhere.
The binary bit flows generated by the encoder network at the UE are fed back to the BS through an uplink feedback channel. The dequantizer module in the decoder network conducts the inverse operation of the quantizer and it generates a 1D compressed tensor. The tensor will be sent to an FCN to restore its original dimension and transform the feature into a sparse representation. Then it is reshaped to a 3D tensor which has the same dimension as the processed channel matrix. After the reshape operation, two JC-ResNets are linked in series to restore the original channel information. Finally, a ``$sigmoid$'' activation layer is used to generate the entries of the recovered channel matrix within $(0, 1)$.
Note that we choose respectively one and two JC-ResNets added to the encoder network and the decoder network for the sake of compromising performance and computational complexity.

\subsection{Joint Convolutional Residual Net}

By connecting a shortcut from the input directly to the output, we construct the JC-ResNet in Fig. \ref{quant_net} as a typical residual network (ResNet) \cite{Resnet} architecture.
Since we commonly use the chain rule to solve the gradient in deep NNs, the gradients returned for shallow network layers are the product of the gradients of backward layers. Considering the gradients are usually less than 1, it leads to the product gradients of the shallow network layers becoming so small that it can be ineffective for network parameter updating.
Therefore, the specific shortcut connection, which shapes the ResNet, can alleviate the issue of gradient dispersion which guarantees better performance.

\begin{figure}[t!]
\centerline{\includegraphics[width=6.0cm,height=2.9cm]{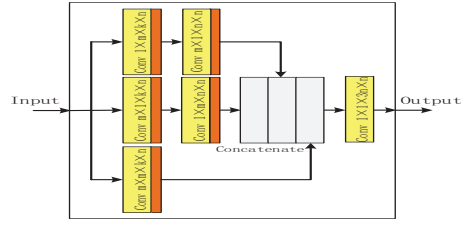}}
\caption{Architecture of a JC block with size $m \times m\times k \times n$.}
\label{joint_conv}
\end{figure}


The setting of the sizes of the convolution kernels is referred to ``RefineNet'' in \cite{CsiNet}. From our experience, the convolution kernel sizes have to grow larger through steps to catch more features. However, to use the residual architecture, the last convolution kernel must have the same output size of the residual block as the input size. Therefore, the sizes of kernels first grow and then decrease. More specifically, in our JC-ResNet, we use 3 joint convolutions (JC), in which the sizes of convolution kernels are $3\times 3 \times 2 \times 8$, $3 \times 3 \times 8 \times 16$ and $3 \times 3 \times 16 \times 2$, respectively.

A JC block with size of $m \times m \times k \times n$ is given in Fig.~\ref{joint_conv}. It includes convolution layers whose kernel sizes directly relate to the JC block. These convolution kernels are actually 4D tensors.
Taking the convolution kernel with size $m \times m \times k \times n$ in Fig. \ref{joint_conv} for instance, the first two dimensions $m \times m$ denote the receptive field size, $k$ denotes the depth of the input tensor, and $n$ denotes the depth of the output tensor.

Considering the fact that the signal path arriving at the receiver at the same time is usually similar in angle and the signal path reached at the same angle is also similar in time delay, the correlation between the horizontal or vertical entries usually is greater than that between the diagonal entries in a real channel sample. Therefore, we propose to use cascaded horizontal or vertical convolutions to catch the horizontal and vertical correlations in steps. In addition, if we use different convolutions in parallel, these convolutions will extract features from different ways. The network will benefit from a ``wide'' structure since it can automatically choose the most effective combination of different features through the learned parameters \cite{Xception}.

Specifically, the input of the joint convolution is spread into 3 flows. The first flow has 2 different convolution operations with the sizes of $1 \times m \times k \times n$ and $m \times 1 \times k \times n$, respectively.
Each convolution operation is added an ``$elu$'' activation layer.
The second flow also has 2 different convolution operations, but with the sizes of $m \times 1 \times k \times n$ and $1 \times m \times k \times n$, respectively. The third flow is a traditional $m \times m$ convolution operation. The first and the second flow are not equal operations owing to the nonlinear activation layers after the convolution operations.
The features extracted by the 3 flows then concatenate together at the last dimension and thus the last depth of the concatenated feature tensor becomes $3n$. Finally, we use a $1 \times 1 \times 3n \times n$ convolution operation, which can not only down-sample the tensor but also make weighted combinations of different features through the channel dimension for a more effective feature expression.

\section{Experiments}
We test our model under the COST2100 indoor channel \cite{COST2100}.
The training, validation and testing sets have 100,000, 30,000 and 20,000 samples, respectively.

All the experiments are conducted on a NVIDIA GTX1080 Ti GPU. In the simulation, we set $N_t=32, N_c=1024$ and ${\tilde{N}}_c=32$. Thus the channel matrix sent to the encoder matrix is ${\bf{H}}^{(a)}_{cp} \in {\mathbb{R}}^{32 \times 32 \times 2}$ and its corresponding complex format ${\bf{H}}^{(a)}_{cc}$, belongs to ${\mathbb{C}}^{32 \times 32}$. The mean squared error (MSE) of the preprocessed channels is chosen as the optimized object during network training:
\begin{equation}\label{eq:mse}
\text{MSE}=\frac{1}{K}\sum_{n=k}^{K-1+k}\sum_{i=1}^{32}\sum_{j=1}^{32}|H_n(i,j) - {\hat{H}}_n(i,j)|^2,
\end{equation}
where $H_n$ is one sample of ${\bf{H}}^{(a)}_{cc}$, $H_n(i,j)$ denotes the entry of the $n$-th sample at position $(i,j)$, ${\hat{H}}_n(i,j)$ denotes the entry of the $n$-th recovered sample at position $(i,j)$, $K$ denotes the training batch size, $k$ denotes the index of the first sample in the batch, and $|\cdot|$ takes the absolute value. The Adam optimizer is used to implement the gradient descent optimization. In order to prevent the explosion of the gradient, we use the gradient clipping measures and the clipping threshold is set to 0.05. Learning rate is set to 0.001. The batch size is set to 200. Another important hyper parameter is the number of quantization bits $B$ for each entry in the compressed feature vector. In the simulation, we set $B$ as 4. All the hyper-parameters are determined with the help of the validation set.


For comparison, we also present the results of CsiNet \cite{CsiNet} and a CS based method TVAL3 \cite{Tval3}. CsiNet-Q($L$) denotes the CsiNet version which is trained without considering the quantization. For testing  CsiNet-Q($L$), each entry of the compressed CSI vector is quantized to $L$ bits. In the simulation, we find that the performance of CsiNet decreases slightly after CSI quantization if $L$ is selected from $\left\{6, 7, 8\right\}$. In addition, we test QCsiNet which only incorporates the operation of quantization in CsiNet and $B$ is set to 4 for fair comparison. The total number of feedback bits is set to 192 for each channel sample in Fig. 5(b).



\begin{figure}[t!]
\centerline{\includegraphics[width=6.3cm,height=2.0cm]{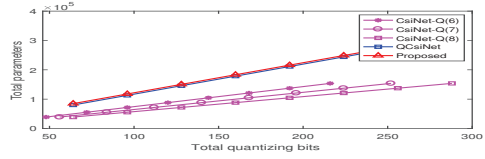}}
\caption{Parameter size of the networks.}
\label{complexity}
\end{figure}



\begin{figure}[t!]
\centerline{\includegraphics[width=6.3cm,height=2.2cm]{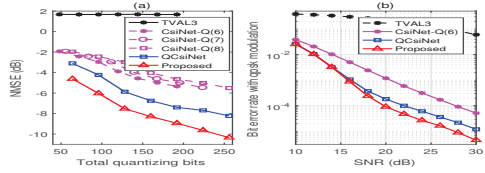}}
\caption{(a) NMSE performance in indoor scenario (b) BER performance in indoor scenario.}
\label{nmse_ber}
\end{figure}


Fig. \ref{complexity} gives the size of parameters of CsiNet, QCsiNet and the proposed NN. From this figure, we conclude that the size of parameters of CsiNet-Q($6$) is about $2/3$ that of ours and QCsiNet has almost the same size of parameters  as the proposed method. Since we quantize each entry of the codeword by 4 bits, the output dimension of the FCN at the encoder is 1.5 times of CsiNet-Q($6$) under the same number of total quantization bits. Therefore, the total parameters is about  1.5 times since it is dominated by the FCNs. As for the computing time, CsiNet-Q($6$), QCsiNet and the proposed NN respectively cost about $0.1$ ms, $0.2$ ms and $0.2$ ms.

Fig. \ref{nmse_ber}(a) compares the normalized mean squared error (NMSE) of our proposed method, CsiNet-Q($6$), CsiNet-Q($7$), CsiNet-Q($8$), QCsiNet and TVAL3. The NMSE is defined as: $\text{NMSE} = 1 / {N_{\text{test}}} \cdot \sum_{k=1}^{N_{\text{test}}}  {{||H_k - {\hat{{H}}}_k||^2}} / {{{||{{H}}_k}||^2}},$
where $N_{\text{test}}$ is the number of channel samples in the testing set, $H_k$ denotes the $k$-th channel sample in the testing set, ${\hat{{H}}}_k$ denotes the $k$-th recovered channel sample and $||\cdot||$ denotes the Frobenius norm. The total number of quantization bits is calculated by the compressed vector dimension multiplied by the number of the quantization bits. The NMSE performance of TVAL3 in Fig. \ref{nmse_ber}(a) is not satisfactory because 200 bits are far from sufficient for TVAL3 to perform well. Fig. \ref{nmse_ber}(b) compares the bit error rate (BER) of these methods. We use the QPSK modulation, the precoding vector is a maximum-rate-transmission (MRT) beamformer designed by using the recovered CSI.
From Fig. \ref{nmse_ber}, QCsiNet achieves a significant gain over the benchmark method, and the design of JC-ResNet further improves the performance by comparing the proposed method with QCsiNet in terms of both NMSE and BER.

\section{Conclusion}

In this paper, we have proposed a more efficient JC-RestNet for CSI feature extraction and recovery. Moreover, we have also provided a method to integrate a quantization operation into the NN, therefore we can optimize the CSI quantization and feedback network at a bit level. Simulation results have shown that the proposed methods can enhance the performance significantly.

\end{document}